# Using learning analytics to provide personalized recommendations for finding peers

Irene-Angelica Chounta[1]

[1] University of Tartu, Estonia
`chounta@ut.ee`

**Abstract.** This work aims to propose a method to support students in finding appropriate peers in collaborative and blended-learning settings. The main goal of this research is to bridge the gap between pedagogical theory and data-driven practice to provide personalized and adaptive guidance to students who engage in computer-supported learning activities. The research hypothesis is that we can use Learning Analytics to model students' cognitive state and to assess whether the student is in the Zone of Proximal Development. Based on this assessment, we can plan how to provide scaffolding based on the principles of Contingent Tutoring and how to form study groups based on the principles of the Zone of Proximal Development.

**Keywords:** learning analytics, personalization, adaptation, student modeling, zone of proximal development, social learning.

## 1    Introduction

The term "*Learning Analytics*" [10] is commonly used to describe the application of computational methods to analyze the learning process and to improve learning outcomes. Even though learning analytics is widely used to support students and instructors in monitoring, mirroring and guiding [5] by providing adaptive and personalized instruction [11], there is little work on using learning analytics to adapt and personalize group formation based on student characteristics. Mainly three approaches are proposed for forming study groups: a) teachers are responsible for setting up groups based on their experience, b) students are free to decide on how to form groups, usually based on their personal relationships or preferences, and c) by applying an algorithmic approach for grouping students based on sets of criteria such as skills, demographics and work styles. However, in most cases these approaches do not take into account the cognitive state of the students. Most importantly, these approaches are disconnected from the teaching or feedback strategies that are followed, from the learning resources students are given and from the learning goals or context in general.

In this work, we propose a holistic, multilevel approach to support collaborative learning activities. This approach combines the use of student models along with established pedagogical theories in order to address students' specific needs on the individ-ual level and to use this information to create meaningful collaborative, learning experiences. In particular, we aim to provide personalized guidance by adapting scaffolding to the students' background knowledge and cognitive state. To monitor and model knowledge and cognitive state, we use computational learning analytics (LA) and machine-learning methods. To maintain the most up-to-date representation of students' knowledge and cognitive state, student models will be dynamically updated during students' practice. Furthermore, we aim to support social aspects of learning by engaging students into collaborative tasks and group discussions with peers. In order to form appropriate study groups – that is, groups that can contribute to collaborative knowledge building – we follow the principles of social development theory and the Zone of Proximal Development (ZPD) [12].

## 2   Background

The Zone of Proximal Development (ZPD) is one of the most popular pedagogical theories. It was introduced by Lev Vygotsky as one of the main themes of the Social Development theory, along with the Social Interaction and the More Knowledgeable Other constructs [4]. The ZPD can be defined as: "*the difference between what a learner can do without help and what he or she can do with help*" [9]. The definition of the ZPD indicates the significance of appropriate assistance in relation to the learning and development process and the importance of choosing suitable peers when forming study groups. Criticism regarding the ZPD usually focuses on the ZPD being an abstract theoretical concept and not an actual, observable and measurable construct [13]. Therefore, it is critical to establish ways for formalizing and operationalizing the ZPD that can contribute towards revealing, understanding and documenting the mechanisms that drive learning and development [8].

There are several approaches to modeling the ZPD. Murray and Arroyo proposed the operationalization of the ZPD as the "state space" between confusion and boredom [8]. They determine when a student is in the ZPD by applying a function of how many hints a student gets correctly in relation to the number of items the student will see. Luckin and du Boulay proposed a design framework (*Ecolab*) that is based on the application of the ZPD for educational software design [6]. The ZPD of a student is assessed using domain knowledge representations, Bayesian Belief Networks (BBN) and two tags, the ability belief and the collaborative support tag. Finally, Chounta et. al. [2, 3] operationalized the ZPD using the concept of the Grey Area, that is the area in which a student model's predictive accuracy is questionable. The rationale is that if the student model cannot predict with acceptable accuracy whether a student is able to carry out a learning task successfully, this may indicate that the student is in her/his ZPD.

## 3   Student modeling for group formation

So far, operationalizations of the ZPD have been used to detect affect such as boredom or frustration [8], to provide learning resources to students [6], or to dynamically adapt tutorial dialogues for students who practice conceptual physics [1]. In this work, we propose the use of the Grey Area operationalization for choosing peers when forming study groups. Based on the Grey Area definition, we can use it as a proxy for assessing when a student is above, below or in her/his ZPD. The Grey Area is individual for each student (just like the ZPD) and it depends on the learning task a student is asked to carry out. On the other hand, the ZPD pinpoints the concept of learning with the assistance of a more knowledgeable other. This means that a low threshold of the ZPD signifies tasks a student can carry out when working alone and the high threshold of the ZPD signifies tasks a student can carry out when working with – or assisted by – a more capable peer. This suggests that in order for learning to potentially take place, a student below her/his ZPD should form a group with a student in or above the ZPD and a student who is in the ZPD should be paired with a student above the ZPD. A student who is already above her/his ZPD can participate in groups of all combinations because this student is already capable of achieving a task on her/his own. This hypothesis has been previously suggested by related research on group formation based on group members heterogeneity with respect to knowledge background [7]. However, it is not yet clear to what extent heterogeneity – or else, distance in terms of background knowledge – supports and facilitates learning. For example, as depicted in Figure 1, if a student below the ZPD (student a) is paired with a student above the ZPD (student c) will she learn more or faster than if she were paired with the student in the ZPD (student b)? One could argue that the larger knowledge "distance" for the first pair (ac) compared to the second pair (ab) will lead to higher learning gains.

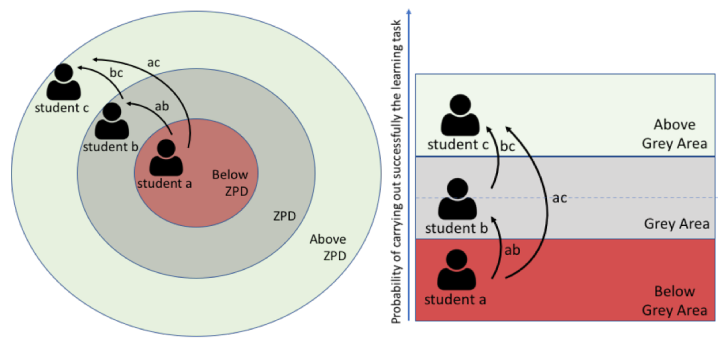

**Fig. 1.** The proposed methodology for forming study groups based on the concept of the ZPD (left) and according to the Grey Area as a proxy of the ZPD (right)

## 4   Discussion

We aim to explore the proposed approach by providing students with collaborative tasks and dynamically assigning group members based on the Grey Area assessment. The research question is to explore how the knowledge "distance" of group members – that is, the difference in terms of their background knowledge – might affect the outcome of their common work and the learning gains both on the individual and group level. The contribution of this work will be twofold: 1. Using machine-learning cognitive models in order to dynamically keep track and assess student's knowledge state and

2. dynamically assigning and adapting study groups based on the participants' knowledge state and with respect to the principles of established pedagogy, namely the ZPD. A potential key broader impact of this work is that it can support complex pedagogical decision-making necessary for providing effective scaffolding. Once the proposed approach has been developed and evaluated, we envision that it can be extended in various contexts, such as online courses, MOOCs and collaborative learning environments.

**Acknowledgements**
This work is supported by the Estonian Research Council grant PSG286.